Positive effects and mechanisms of simulated lunar low-magnetic environment on earthworm-improved lunar soil simulant as a cultivation substrate

Sihan Hou[#,a,b,c], Zhongfu Wang[#,a,b,c], Yuting Zhu [a,b,c], Hong Liu[*,a,b,c], Jiajie Feng[*,a,b,c],

[a]*Institute of Environmental Biology and Life Support Technology, School of Biological Science and Medical Engineering, Beihang University, Beijing, 100191, China*

[b]*Beijing Advanced Innovation Center for Biomedical Engineering, Beihang University, Beijing, 100191, China*

[c]*International Joint Research Center of Aerospace Biotechnology & Medical Engineering, Beihang University, Beijing, 100191, China*

*Corresponding authors (fengjiajie@buaa.edu.cn; lh64@buaa.edu.cn)

[#]These author contributed equally to this work.

# Abstract

With the advancement of crewed deep-space exploration missions, Bioregenerative Life Support Systems (BLSS) for lunar-surface bases has become crucial, which would face the stresses from lunar *in situ* environmental factors. Compared to microgravity and long-term low-dose ionizing radiation, the low-magnetic field is difficult to simulate in Earth-based experiments. However, it may not necessarily act as a stress for BLSS biological components, so elucidating it may significantly simplify the subsequent research. Earthworms, known as "soil scavengers", have been proved as efficient in improving lunar soil simulant as a plant cultivation substrate and the degradation-recycling of inedible lignocellulosic plant residues by our previous studies. This study applied earthworms to a mixed substrate of lunar soil simulant and the organic solid waste (from our "Lunar Palace 365" crewed long-term closed BLSS experiment) under three magnetic conditions: lunar low-magnetic (5 nT), Earth-magnetic (25–65 μT), and high-magnetic (7,000 μT). Results showed that stronger magnetic fields significantly increased the earthworms' oxidative stress levels (MDA) and impaired neurotransmitter levels ($Ca^{2+}/Mg^{2+}$-ATPase and acetylcholine). Consistently, weaker magnetic fields showed significantly better earthworm performance in improving the substrate's cultivability, including pH (more neutral), available nutrients, humus content and wheat seedling rate. Microbial community analyses revealed more mechanisms: the higher earthworm-gut fungal Shannon index (α-diversity) under high-magnetic field verified the earthworm's impaired grinding/digesting function; for earthworm gut molecular ecological networks under stronger magnetic fields, the higher proportion of positive correlations suggests slower microbial cooperations (e.g., lignocellulose degradation task); the decreased Network Size, Average Path Length and Modularity indicated a substantial disruption of microbial interactions.

This study disproves the stressful role of lunar (or space) low-magnetic field on the earthworm-lunar soil-waste system in BLSS, providing important reference for further deep-space BLSS research.

## 1. Introduction

The advancement of deep space exploration will significantly expand human living space while driving scientific research and economic development. Crewed exploration is becoming an inevitable trend in future deep space missions. As the first destination for human deep space exploration, the establishment of crewed lunar bases has been placed on the agenda. To ensure the long-term residence of astronauts on the lunar surface, transporting supplies from Earth incurs prohibitively high costs. For instance, the United States' Space Launch System (SLS) can deliver 70–130 tons of materials into Earth-Moon transfer orbit per launch, yet the cost reaches approximately $11.8 billion (GAO, 2023). In contrast, the Bioregenerative Life Support System (BLSS), as an efficient and sustainable life support technology, has garnered widespread attention (Fu et al., 2016). BLSS integrates ecosystem technology with engineering control, enabling self-sufficiency in oxygen, water, and food within a closed-loop ecological system through plants, microbes, and other biological processes. This drastically reduces the cost of transporting life-support materials from Earth while enhancing the sustainability and stability of lunar bases. In the "Lunar Palace 365" Earth-based experiment conducted by our team in 2018, BLSS successfully operated a closed ecosystem for over a year, ensuring the long-term health of volunteers and demonstrating its potential for prolonged space missions (Fu et al., 2021).

The efficient utilization of in situ lunar resources will further enhance the feasibility of lunar BLSS. In terms of life support, plant cultivation serves as the core functional unit of BLSS, generating fresh $O_2$, food, and water (Zabel et al., 2016). Moreover, interactive activities such as crop cultivation promote the release of neurotransmitters like dopamine and serotonin while reducing glucocorticoid secretion, thereby safeguarding astronaut mental health (Meng et al., 2020). Lunar soil simulant holds potential as a hydroponic substrate for BLSS crop production (Burke and Poulet, 2014). Studies indicate that plants can grow in genuine lunar regolith, albeit with stunted development and severe phenotypic stress, making it unsuitable as an ideal cultivation substrate

(Paul et al., 2022). Consequently, researchers have explored various improvement methods, including the use of plant/animal residues (Wamelink et al., 2021), Bacillus spp. (Lytvynenko et al., 2006), and livestock manure (Caporale et al., 2022), all of which significantly enhance plant growth. However, these approaches have yet to integrate lunar soil improvement with BLSS operations. During BLSS operation, substantial solid waste is generated, primarily consisting of human feces and partially decomposed plant residues, both rich in microbes. We previously mixed lunar soil simulant with solid waste from the "Lunar Palace 365" experiment at varying ratios, conducting aerobic co-fermentation before applying it to wheat cultivation (Yao et al., 2021). Results showed that adding small amounts of solid waste to the simulant enabled wheat seedlings to reach 60% of the height achieved in vermiculite cultivation, demonstrating significant improvement potential for lunar regolith.

Efficient solid waste treatment is also critical for BLSS stability. Waste management not only ensures detoxification but also directly impacts resource recycling within BLSS (Jin et al., 2021). Conventional methods include physical, chemical, and biological treatments. While physical and chemical approaches enable rapid processing, they often rely on non-renewable lunar materials and risk incomplete treatment or secondary pollution. In contrast, biological treatment—owing to its efficiency and environmental friendliness—has emerged as a core technology, including aerobic fermentation (Hendrickx et al., 2006), anaerobic digestion (Albiol et al., 2002), tilapia aquaculture (Gonzales and Brown, 2007), and microalgae cultivation (Verstraete et al., 2016). However, BLSS organic waste mainly comprises inedible plant residues (~90% dry weight), dominated by lignocellulose, which is recalcitrant to degradation. Thus, coupling lunar soil cultivability improvement with organic waste recycling remains a key challenge for lunar BLSS.

Through literature review and experimentation, we selected earthworms as an ecological solution to this dual challenge. As ubiquitous soil organisms, earthworms are widely used in waste treatment and soil improvement (Domínguez, 2023). By ingesting and excreting organic waste, they convert it into nutrient-rich vermicast, significantly enhancing soil physicochemical properties (Aira et al., 2007). Additionally, earthworm activity boosts microbial diversity and activity, improving waste processing efficiency by transforming waste into humus and organic fertilizers that enhance plant growth (Gómez-Brandón et al., 2008). Donato Romano et al. inoculated Eisenia fetida into LHS-1 lunar soil simulant for 60 days, confirming their survival, feeding, and cocoon production

(Romano et al., 2023). Our preliminary experiments introduced organic waste and E. fetida into spring wheat cultivation in simulant, observing significant improvements in edible biomass, 1,000-grain weight, plant height, leaf length, chlorophyll content, net photosynthetic rate, relative water content (RWC), membrane stability index (MSI), and root malondialdehyde (MDA, reflecting stress levels) compared to pure simulant. Plant height reached 82% of that in vermiculite with nutrient solution (Wang et al., 2025), preliminarily validating earthworms' potential for BLSS waste treatment and lunar soil improvement.

A well-designed lunar soil improvement-waste conversion coupled unit could greatly enhance BLSS efficiency and robustness. However, lunar BLSS and its biological components face another major challenge: the Moon's environmental conditions—even within enclosed habitats—differ drastically from Earth's, including low gravity (~1/6 g) and low magnetic fields, whose stressful effects on BLSS remain unclear. Thus, preemptive lunar life science research is essential to ensure astronaut survival and health. We previously found that simulated microgravity significantly inhibits lignocellulose waste degradation by restricting and fragmenting microbial communities, weakening synergies while increasing antifungal metabolites and competition (Liao et al., 2025). Therefore, earthworm applications in lunar BLSS require validation under space or simulated space conditions. Yet, space-based earthworm experiments remain virtually unexplored. Hence, this study focuses on lunar low magnetic conditions, investigating earthworm-mediated improvement of lunar soil simulant-BLSS waste mixed substrates and its underlying mechanisms.

In this study, we blended 30% BLSS solid waste (dry w/w) with lunar soil simulant to form a mixed substrate, then subjected it to earthworm-mediated improvement for 10 days under simulated lunar low magnetic, Earth magnetic, and permanent magnet-generated high magnetic conditions. Controls without earthworms were maintained under Earth magnetic conditions. We comprehensively analyzed earthworm physiology, substrate improvement effects, microbial community composition (substrate and gut), and their interaction models. Results indicate that lunar low magnetic conditions exert positive effects on earthworms and their substrate improvement process. Earthworms not only maintained healthy physiological status but also enhanced microbial functionality, accelerated humus decomposition, and optimized substrate properties for crop cultivation. For lunar BLSS, this study alleviates concerns about low magnetic effects on cultivation substrates, affirming earthworms' potential in lunar soil improvement and waste treatment.

## 2. Materials and methods

### 2.1 Mixed substrate, earthworms and magnetic field environments

The organic solid waste used in this study was derived from our "Lunar Palace 365" experiment. To ensure representativeness, we mixed solid waste samples collected on July 4, 2017 (early phase of the experiment) and May 15, 2018 (late phase). The lunar soil simulant used was CUG-1B, provided by Prof. Long Xiao's research group at China University of Geosciences (Wuhan). The solid waste and lunar soil simulant were mixed at a 3:7 ratio, adjusted to 70% humidity, and placed in plastic bottles with ventilation holes.

*Eisenia fetida*, a commonly used laboratory model species, was obtained from a worm breeding farm (Changzhou, Jiangsu, China). Prior to the experiment, the earthworms were acclimatized for two weeks to adapt to the experimental conditions. Mature adult earthworms ($0.5 \pm 0.39$ g) were selected for the study and underwent thorough disinfection to ensure their effectiveness in improving soil properties and preventing the introduction of pathogenic pests into the cultivation substrate.

All magnetic field treatments utilized a standardized cylindrical growth pot (65 cm diameter × 200 cm height, total volume ≈663,662 mL) constructed from polypropylene. The container was specifically designed to maintain consistent physical conditions across all magnetic field environments while accommodating the different field generation apparatus.

Three magnetic field conditions were designed for comparative experiments, with each magnetic field treatment having 3 identical growth pots as replicates: (1) Low magnetic field group: growth pots were housed within a four-layer permalloy magnetic shielding device (Fig. 1a). Residual magnetic fields were measured at nine representative locations using a calibrated fluxgate magnetometer (Mag03 sensor, resolution 0.1 nT, bandwidth 3 kHz, range ±100 µT at 25°C). All positions maintained field levels below 5 nT (Fig. 1b). (2) Earth magnetic field control group: growth pots was placed on non-magnetic wooden tables (residual magnetism <5 nT) in an electromagnetic shielding chamber (background noise <10 nT), maintaining $25.3 \pm 0.8$ µT ambient field. Field uniformity was verified using the same magnetometer. (3) High magnetic field group: twenty N52-grade NdFeB permanent magnets (30×20×5 mm) were arranged in a Halbach array surrounding growth pots (Fig. 1c). Field uniformity was verified at multiple points (Fig. 1d), with precise spacing adjustments achieving 7,000 ± 250 μT central field strength (gradient <50 μT/cm).

Ten earthworms were placed in each substrate for a 10-day observation period. Post-experiment, physiological indices of earthworms and physicochemical properties of the substrate were analyzed.

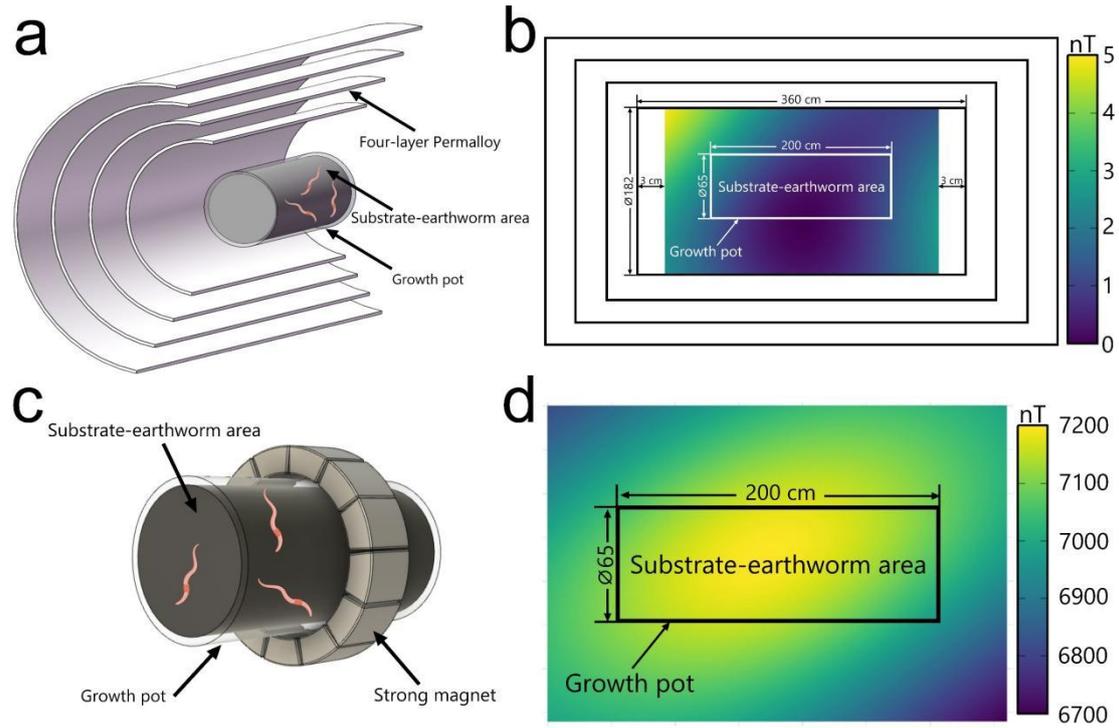

**Fig. 1.** Schematic diagrams of the four-layer permalloy device and the earthworm growth pot's setup. (a) The placement of the earthworm growth pot inside the four-layer permalloy device; (b) Result of the residual magnetism test in the magnetic shielding device; (c) The placement of the earthworm growth pot inside the strong magnet device; (d) Result of magnetic test in the strong magnet device.

## 2.2 Earthworm physiological indices

Histopathological evaluation (HE) staining was performed following standard histopathological methods. Intestinal tissue morphology and inflammation levels were observed and imaged at 200× magnification. Paraffin sections were dehydrated with xylene and absolute ethanol, stained with hematoxylin and eosin, and mounted with neutral resin for microscopic examination and image analysis.

After gut content removal, intestinal tissue samples were homogenized in saline and analyzed for SOD and MDA content using commercial kits (Nanjing Jiancheng Bioengineering Institute,

China). Whole brain tissue from each earthworm was homogenized in saline, and acetylcholine levels, $Ca^{2+}$-ATPase, and $Na^+$-ATPase activities were quantified using commercial kits (Nanjing Jiancheng Bioengineering Institute, China).

## 2.3 Substrate physicochemical properties

Lunar soil simulant samples (25 ± 2.5 g) were extracted using undisturbed soil samplers. Extract pH and electrical conductivity (EC) were measured with pH (FE28, METTLER TOLEDO, Switzerland) and EC meters (FE38, METTLER TOLEDO). Soil organic matter (SOM) was determined via potassium dichromate titration. Humus content was measured after extraction with 0.1 mol/L sodium pyrophosphate and 0.1 mol/L sodium hydroxide, followed by dichromate oxidation. Total carbon (C) and nitrogen (N) were analyzed using an elemental analyzer (UNICUBE, Elementar, Germany). Hydraulic conductivity (HC) was determined via constant-head permeability tests and Darcy's law.

## 2.4 Microbial sequencing of earthworm gut and substrate

Comprehensive information about the bacterial community composition in cultivation substrate was obtained through 16S rRNA high-throughput sequencing. DNA extraction, PCR amplification, and sequencing of cultivation substrate microbial communities were performed by Biomarker Tech. Corp., Beijing, China. Briefly, DNA was extracted with the TGuide S96 Magnetic Soil/Stool DNA Kit (Tiangen Biotech (Beijing) Co., Ltd., China) according to the manufacturer's instruction. DNA concentration was measured with the Qubit dsDNA HS Assay Kit and Qubit 4.0 Fluorometer (Invitrogen, Thermo Fisher Scientific, Eugene, Oregon, USA). The 338F: 5'-ACTCCTACGGGAGGCAGCA-3' and 806R: 5'-GGACTACHVGGGTWTCTAAT-3' universal primer set was used to amplify the V3-V4 region of 16S rRNA gene. Both the forward and reverse primers were tailed with sample-specific Illumina index sequences. The PCR was performed in a total reaction volume of 10 μl: DNA template 5-50 ng, *Vn F (10 μM) 0.3 μl, *Vn R (10 μM) 0.3 μl, KOD FX Neo Buffer 5 μl, dNTP (2 mM each) 2 μl, KOD FX Neo 0.2 μl, and dd$H_2O$ up to 10 μl. Vn F and Vn R were selected according to the amplification area. The initial denaturation at 95°C for 5 min was followed by 25 cycles of denaturation at 95°C for 30 s, annealing at 50°C for 30 s and extension at 72°C for 40 s, and a final step at 72°C for 7 min. PCR amplicons were purified with Agencourt AMPure XP Beads (Beckman Coulter, Indianapolis, IN, USA) and quantified using the Qubit dsDNA HS Assay Kit and Qubit 4.0 Fluorometer (Invitrogen, Thermo Fisher Scientific). The quantified amplicons were then pooled together in equal amounts as the library. The constructed library was sequenced using Illumina NovaSeq 6000 (Illumina, Santiago, CA, USA).

Downstream sequencing analysis was performed on BMK Cloud (Biomarker Technologies Co., Ltd., Beijing, China). Raw sequences were first processed using Trimmomatic and FLASH, with a moving window of 50-bp and a quality threshold score of 30. Singletons were then removed. Next, high-resolution amplicon sequence variants (ASVs) were identified from the reads using DADA2 (version 2020.06). Lastly, a representative sequence of each ASV was annotated through SILVA ribosomal RNA gene database (version 132) with a confidence score of 0.7.

## 2.5 Molecular ecological network analysis

With 16S rRNA gene (bacteria) ASVs, ITS ASVs and environmental factors pooled together as the input, phylogenetic molecular ecological networks (pMENs) were constructed based on random matrix theory (RMT)-based network (Deng et al., 2012b). The threshold of similarity coefficients (r values of the Spearman's rho correlation) for network construction was automatically determined when the nearest-neighbor spacing distribution of eigenvalues transitioned from Gaussian orthogonal ensemble to Poisson distributions (Deng et al., 2012b). Random networks corresponding to all pMENs were constructed using Maslov-Sneppen procedure with the same network size and average link number to verify the system-specificity, sensitivity and robustness of the empirical networks(Maslov and Sneppen, 2002). Network graphs were visualized with Cytoscape 3.8 software.

## 2.6 Process of microbial community assembly

To explore the structure of bacterial and fungal community assembly processes by deterministic or stochastic processes (Stegen et al., 2012), the β-nearest taxon index (β-NTI; was calculated using the R package "picante" (version 1.8.2). The β-NTI value, calculated using null-model expectations with consideration for phylogenetic distance, provides insight into the turnover of microbial communities. A null distribution of Beta Mean Nearest Taxon Distance (β-MNTD) is performed by randomizing OTUs across the phylogeny and recalculating β-MNTD 999 times (Stegen et al., 2012). B-NTI quantifies the number of standard deviations that the observed β-MNTD is from the mean of the null distribution.

## 2.7 Statistical analysis

Shannon index (α-diversity) was calculated and displayed with the package "ieggr" (version 4.17) using R software (version 4.3.3). Principal coordinate analysis (PCoA) was calculated and displayed with the package "ape" (version 5.0) using R software. Generalized linear models were constructed by "glm" function in R to investigate relationships among relative abundance of key-microbes, and wheat agronomical parameters. Redundancy analysis (RDA) among microbial communities and environmental factors was performed using BMKCloud (www.biocloud.net, Biomarker Tech. Corp.). Analysis of variance (ANOVA) and post-hoc Fisher's Least Significant Difference (LSD) test with Bonferroni P-adjust determining the difference among cultivation groups were calculated with the package "agricolae" (version 1.3-7) using R software. All experiments were performed 3 or more times and experimental results were analyzed for statistical significance using 2-tail test.

# 3. Results

## 3.1 Effects of magnetic fields on physiological and intestinal functions of earthworms

To assess the effects of different magnetic environments on earthworm physiology and intestinal function, we measured oxidative stress indicators and conducted hematoxylin-eosin (HE) staining to analyze intestinal tissue structure and functional status. The results revealed that the high magnetic environment significantly increased malondialdehyde (MDA) levels (Fig. 2a), indicating induced oxidative stress in earthworms. Concurrently, the activity of $Ca^{2+}/Mg^{2+}$-ATPase (Fig. 2b) and acetylcholine levels (Fig. 2c) were markedly reduced. Since acetylcholine release is triggered by calcium ion channels, this suggests that the high magnetic field likely exerts a stronger Lorentz force on divalent ions, impairing downstream neurotransmitter release and thus compromising neural function. In contrast, $Na^+/K^+$-ATPase (monovalent ion channel) activity remained unaffected (Fig. 2d), further supporting this mechanism.

HE staining (Fig. 2e) demonstrated that in both the Earth magnetic and low magnetic groups, the circular and longitudinal muscle layers exhibited well-defined structural boundaries, uniform coloration, and a plump, compact morphology without pathological changes, suggesting normal muscle development and peristaltic function. In contrast, the high magnetic group displayed shrunken, unevenly colored muscle layers with localized thinning and even signs of detachment, likely due to impaired muscle contraction caused by magnetic interference with $Ca^{2+}$ channel function (e.g., reduced $Ca^{2+}$-ATPase activity). Additionally, the intestines of earthworms in the low magnetic and Earth magnetic groups appeared rounded, intact, and tightly structured, with neatly distributed mucosal folds indicating high surface area and absorption efficiency. Conversely, the high magnetic group exhibited brittle, thin, and loose intestinal walls with smoother mucosal folds, suggesting reduced absorptive function under high magnetic conditions.

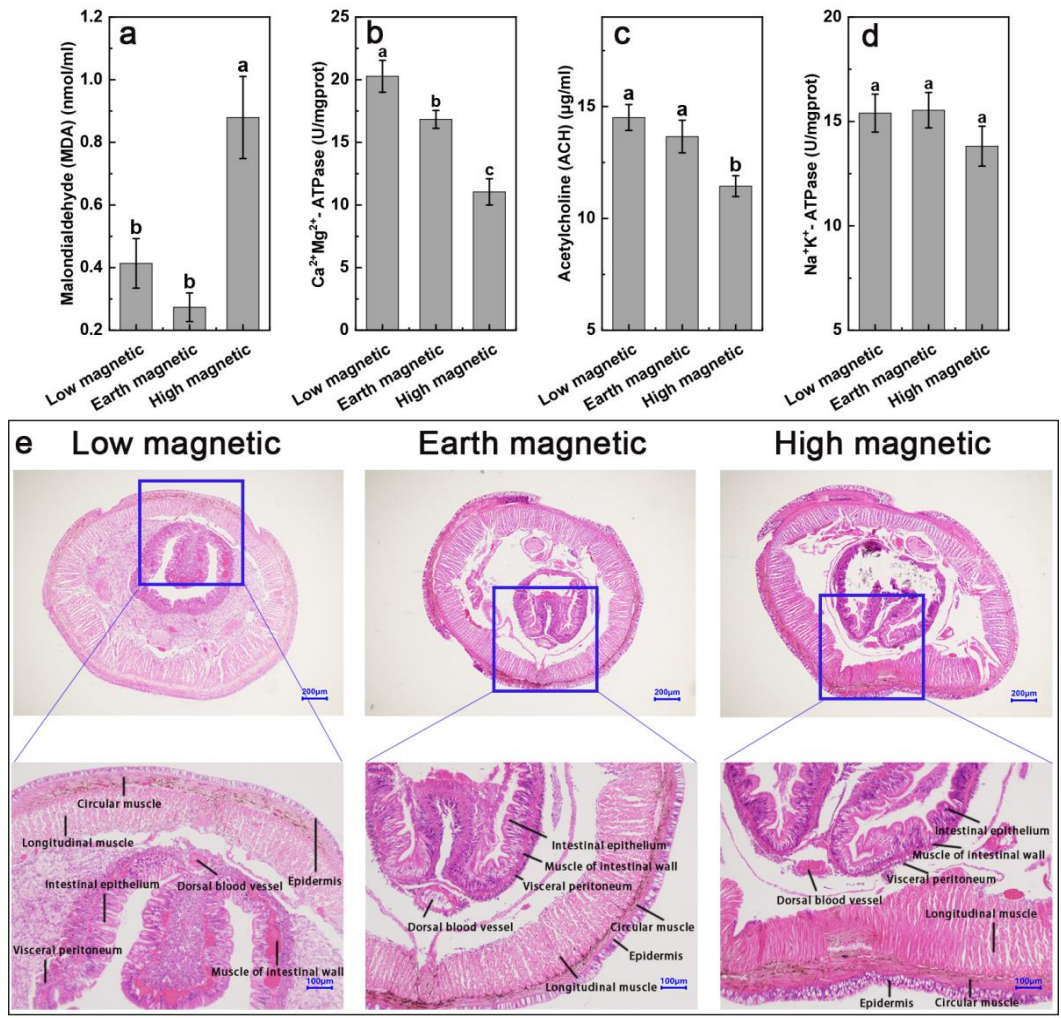

**Fig 2.** Physiological indicators of earthworms in different magnetic field environments, including (a) malondialdehyde (MDA), (b) $Ca^{2+}Mg^{2+}$ - ATPase, (c) acetylcholine (ACH), (d) $Na^+K^+$ - ATPase, and (e) hematoxylin- and eosin- (H & E) stained cross-sectional image of the earthworms' intestine. Letters (e.g. a and b) above the bars indicate the result of ANOVA's post-hoc LSD test identifying significant differences among the labeled bars.

### 3.2 Substrate physicochemical properties related to plant cultivation

To evaluate the improvement effects of earthworms on solid waste-lunar soil simulant mixed substrates under different magnetic environments, we measured substrate salinity-alkalinity indicators (pH, electrical conductivity, salinity) and chemical composition indicators (total carbon content (C), total nitrogen content (N), available nitrogen, available phosphorus, available potassium, and humus content). The results showed that the control group (incubated for the same duration as other treatment groups) had significantly higher pH (Fig. 3a), electrical conductivity (Fig. S1a) and salinity (Fig. S1b) than other groups, while the low magnetic and Earth magnetic groups showed lower values, indicating that earthworm activity helped alleviate substrate alkalinity and salinization.

Regarding substrate maturation, chemical composition analysis showed that the high magnetic group had significantly higher total carbon content (Fig. 3b) and C/N ratio (Fig. S1c) than the low and Earth magnetic groups, suggesting that the high magnetic field may hinder substrate maturation. Humus content (Fig. 3c) was significantly lower than in the low magnetic group. As humus is a stable product of microbial (e.g., fungi, *Actinobacteria*) decomposition of plant residues, this indicates that the high magnetic environment may strongly inhibit earthworm activity, hyphal growth, or lignin-degrading enzyme secretion, thereby impeding organic matter degradation in the substrate and likely resulting in insufficient maturation.

In terms of plant-available nutrients, the low magnetic group showed significantly higher available nitrogen (Fig. 3d) and potassium (Fig. 3e) contents than other groups. Available nitrogen and potassium mainly originate from microbial nitrogen fixation and metabolic activities, suggesting that under low magnetic conditions, earthworms may increase mineral fragmentation efficiency while enhancing microbial conversion efficiency of plant-available nutrients. Although phosphorus content (Fig. S1d) was unaffected by the magnetic field, the earthworm-treated groups still showed significantly higher levels than the no-earthworm control. In summary, different nutrients responded differently to magnetic fields: carbon degradation and available potassium were only sensitive to high magnetic fields, while available nitrogen and humus showed significant improvements only under low magnetic conditions (Fig. 3).

To directly verify the effects of improved substrates on plant cultivation, we conducted a 7-day wheat germination experiment using substrates from each group. The results showed that in low and Earth magnetic environments, earthworm-improved substrates significantly increased wheat germination rates, while the high magnetic group showed poor germination performance, with no significant difference from the control (Fig. 3f), indicating unsatisfactory substrate improvement by earthworms under high magnetic fields. For wheat seedling height at 7 days (calculated only for germinated individuals, Fig. S1e), there were no significant differences among magnetic field-earthworm groups, but all were significantly higher than the control.

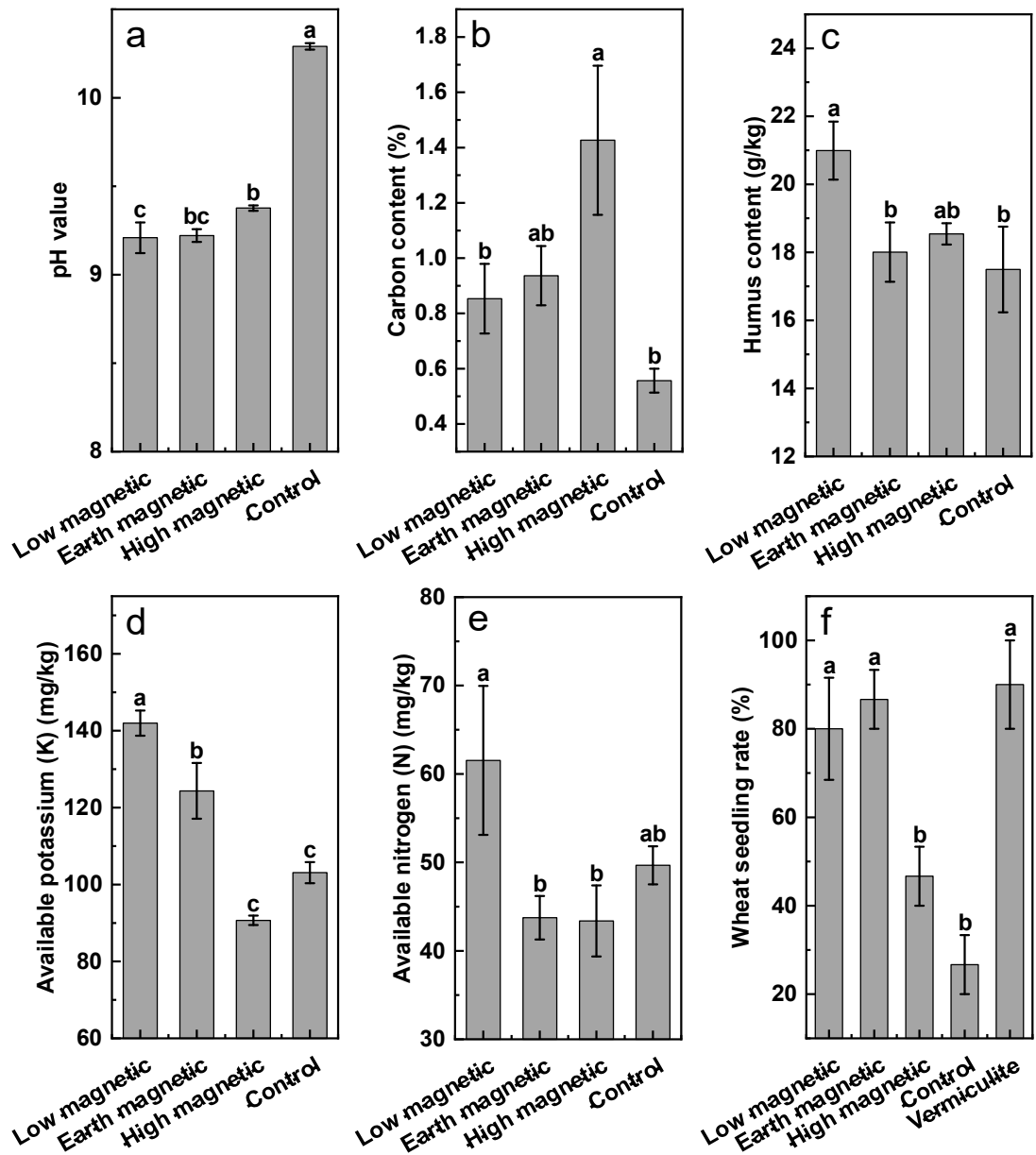

**Fig 3.** The substrates' physical and chemical properties related to crop cultivation, including (a) pH, (b) total carbon content (C), (c) available nitrogen, (d) available potassium, (e) humus content, and (f) germination rate of wheat seedlings after 7 days. Letters (e.g. a and b) above the bars indicate the result of ANOVA's post-hoc LSD test identifying significant differences among the labeled bars.

## 3.3 Microbial diversity, composition, and functional profiling in substrate and earthworm gut

To explore the improvement mechanism from the aspect of the microbial communities, 36 samples (= 3 treatment groups ×12 biological replicates) of earthworm gut contents and 33 substrate samples (= 3 treatment groups × 9 biological replicates + 6 control samples) underwent amplicon sequencing (removed five low-quality 16S rRNA samples), and generated an average of 58,188 non-chimeric amplicon sequences for bacterial 16S rRNA gene and 69,743 for fungal ITS region, which were clustered into 24,199 bacterial ASVs and 16,760 fungal ASVs. By the annotation, bacteria were classified into 1,594 species, and fungi were classified into 1,781 species.

In α-diversity analysis, we calculated the Chao1 index to estimate species richness (i.e., the number of species in the community) and the Shannon index to consider more weighted species richness and evenness. For bacterial communities, the results showed that the species richness (Chao1 index) of substrate bacterial communities was significantly higher than that of earthworm gut and control groups (Fig. 4a), indicating that earthworm activities (burrowing and feeding) might create more microhabitats and niches in the substrate, while the strong selective pressures of low oxygen, high pH, and digestive enzymes in the earthworm gut reduced the niches. However, there was no significant difference in bacterial species evenness (Shannon index) between the substrate and earthworm gut (Fig. 4b), suggesting that bacteria in these newly added niches had not yet proliferated significantly. Magnetic treatments did not significantly alter bacterial diversity (Fig. 4a & 4b).

For fungal communities, species richness (Chao1 index) showed no significant differences among treatment groups (Fig. 4c). However, species evenness (Shannon index) in the earthworm gut and control groups was significantly higher than in the substrate group (Fig. 4d), indicating that earthworm activities reduced fungal diversity in the substrate, likely due to the destructive physical grinding of relatively larger fungal structures by the earthworm gizzard and the killing of fungi by phenol oxidase and lysozyme secreted in the gut. Additionally, the Shannon index of substrate fungi was significantly higher under high magnetic conditions (Fig. 4d), possibly because the high magnetic field interfered with calcium ion signaling in gizzard muscle cells (e.g., by inhibiting $Ca^{2+}$-$Mg^{2+}$-ATPase, Fig. 2b), leading to reduced efficiency of physical grinding of fungi.

To identify differences in community structure among groups, we performed detailed comparisons using distance matrix-based principal coordinate analysis (PCoA). The results showed that the effects of magnetic treatments were not significant. At the bacterial level, the bacterial communities of the control group, earthworm gut, and improved substrate each showed distinct differences (Fig. 4e), indicating that earthworm activities significantly and directionally converged to alter the bacterial community structure in the substrate. At the fungal level, the fungal community of the earthworm-improved substrate still differed significantly from that of the gut, but the gut community structure was surprisingly similar to that of the control group (Fig. 4f), possibly because in situ fungi were extremely rare in the earthworm gut microbial community, and almost all measured fungi originated from ingested food.

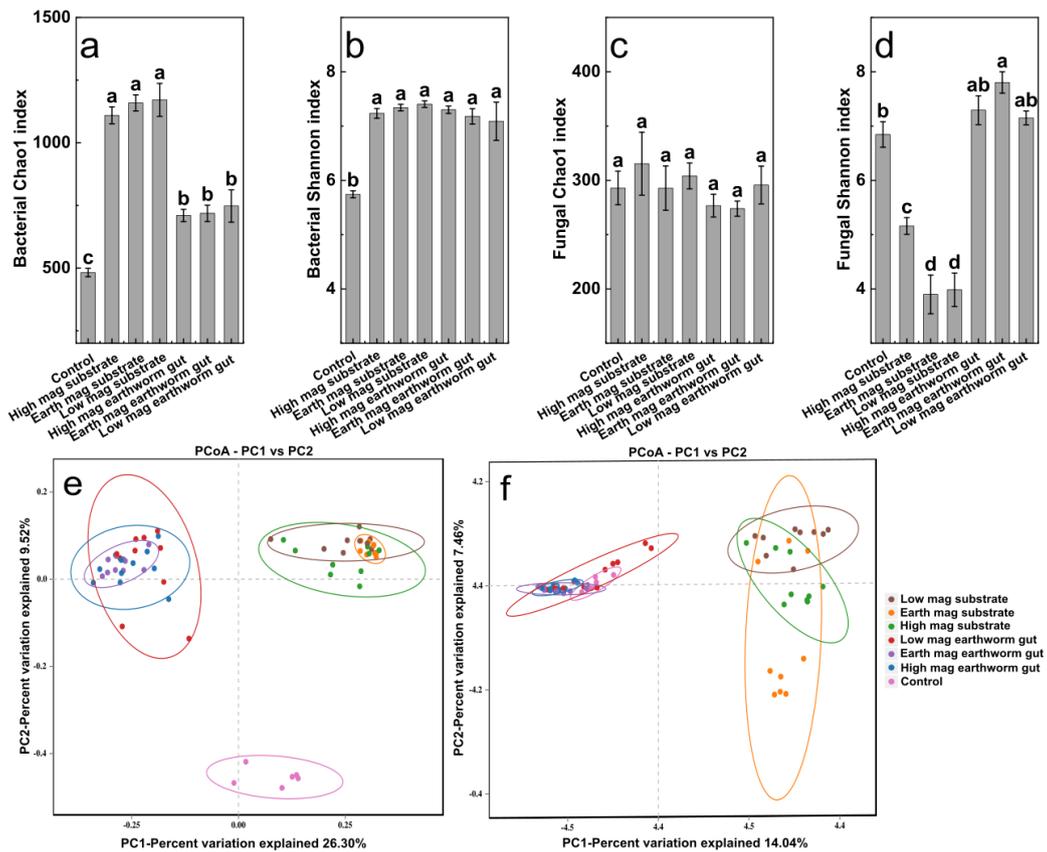

**Fig 4.** Microbial community diversity analyses, including α-diversity and β-diversity analyses. The α-diversity analyses includes: (a) Chao1 index of bacterial communities, depicting species richness, (b) Shannon index of bacterial communities, depicting both species richness and evenness, (c) Chao1 index of fungal communities, and (d) Shannon index of fungal communities. Letters (e.g. a and b) above the bars indicate the result of ANOVA's post-hoc LSD test identifying significant differences among the labeled bars. The β-diversity analyeis includes: (e) Principal Coordinate

Analysis (PCoA) plot of bacterial communities, and (f) PCoA plot of fungal communities, based on Bray-Curtis distances.

To further investigate the effects of earthworms on substrate microbes under different magnetic environments, we conducted functional prediction analyses, with particularly noteworthy findings regarding fungal communities. In the functional structure of fungal communities, the earthworm gut exhibited similarity to the control group (Fig. S2a), suggesting that gut fungi likely originated from fresh substrate, showing parallels in both structural composition (Fig. 4f) and functional units. Magnetic fields demonstrated significant perturbations on fungal communities. High magnetic treatment enhanced plant pathogen-related functional attributes (Plant Pathogen) in both substrate and earthworm gut samples, while simultaneously suppressing saprotrophic functions (Undefined Saprotroph) - specifically inhibiting organic matter decomposition capacity (Fig. S2a). These alterations represent unfavorable directions for cultivation substrate improvement. Bugbase phenotype prediction results further revealed that earthworms could reduce potentially pathogenic microorganisms (Potentially_Pathogenic) (Fig. S2b). Concurrently, earthworms significantly increased biofilm-forming bacterial functions (Forms_Biofilms). The extracellular polymeric substance (EPS) matrices formed by these microbial biofilms can promote substrate aggregate structure formation, representing a crucial element in cultivation substrate improvement.

### 3.4 Microbial network analysis and community assembly mechanisms

To delve into functional interactions within earthworm gut and substrate microbial communities, we constructed seven molecular ecological networks (pMENs) with bacteria and fungi pooled together (Fig. 5). As anticipated, each network exhibited topological properties of small world, scale-free and modularity, and were significantly different from randomly generated networks (Table S1), indicating a qualified representativeness for complex system's networks. Zi-Pi parameters and modular structures further revealed the distribution characteristics of hub nodes under different magnetic conditions (Fig. S3).

The microbial molecular ecological networks (Fig. 5) demonstrated that compared to substrates, the microbial networks in earthworm guts showed more pronounced trends of being shaped by magnetic fields, likely due to stronger magnetic effects on animals. The proportion of positive correlations in gut networks increased significantly with field strength (Fig. 5), suggesting that high magnetic fields may slow down microbial cooperative tasks (e.g., lignocellulose degradation), leaving them incomplete. Network Size, Average Path Length and Modularity of gut networks all decreased progressively with increasing field strength (Table S1), indicating substantial disruption of microbial interactions in the gut by magnetic fields. The networks under low magnetic conditions exhibited significantly greater size, modularity, and number of hubs compared to both Earth magnetic and high magnetic groups (Fig. 5; Table S1), suggesting that low magnetic fields may potentially activate functional capabilities of earthworm gut microbiota.

Substrate microbial networks exhibited notably different structures and compositions from gut networks, with less obvious magnetic shaping trends (Fig. S3). However, the proportion of positive correlations in substrate networks also increased significantly with field strength (Fig. S3). Both low magnetic and Earth magnetic groups showed lower proportions of positive correlations between nodes (48.33% and 49.57% respectively) and fewer hub nodes (only 1 each). In contrast, the high magnetic substrate network resembled the unimproved control network more closely, with positive correlation proportions of 54.35% and 53.77%, and hub node numbers of 5 and 3, respectively. The dominant bacterial phyla and their proportions in substrate networks under Earth, low and high magnetic fields were similar (Figure S3), primarily *Proteobacteria* (~50%), *Bacteroidota* (~25%) and *Verrucomicrobiota* (~7%). The unprocessed control substrate displayed different dominant phyla, particularly with the emergence and higher proportion (~14%) of fungi (*Ascomycota*), demonstrating earthworms' strong shaping capacity on functional bacterial phyla and significant inhibitory effects on fungi across various magnetic conditions.

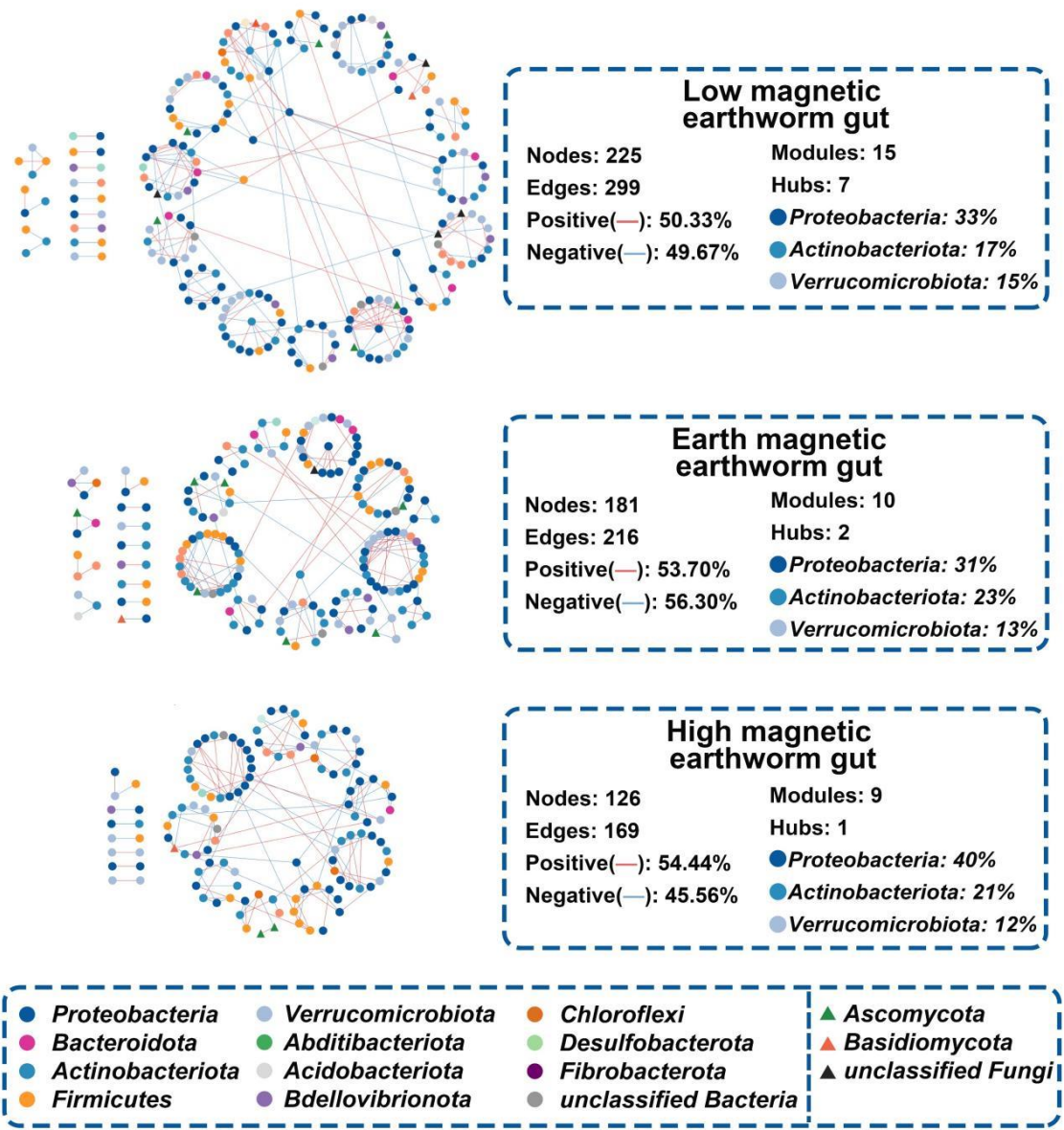

**Fig 5.** Earthworm gut microbial communities' interactions under different magnetic field conditions. In the molecular ecological network, circular nodes represent bacterial species, while triangle nodes represent fungal ones; the nodes' colors distinguish the phylum-level taxonomy; red links indicate positive correlations between two nodes, while blue links indicate negative correlations.

Community assembly refers to the ecological processes that govern the formation and organization of microbial communities. Among these processes, deterministic assembly (including heterogeneous selection and homogeneous selection) typically increases with environmental stress, while stochastic assembly (including homogenizing dispersal and drift) decreases with environmental stress. It should be noted that microbial dispersal can be either stochastic or deterministic, and quantitative evidence supporting the inverse relationship between drift and environmental stress remains limited. To assess the relative contributions of various assembly processes, we employed the phylogenetic-bin-based null model analysis for inferring community assembly mechanisms (iCAMP).

In bacterial communities, homogeneous selection predominated (51.81 ± 2.37%; Fig. 6a) and showed significant correlations with both earthworm gut physiology and substrate physicochemical properties (Fig. 6b), indicating that the selective pressures created by earthworm digestion shaped bacterial communities with relatively uniform niche distributions. Although no significant differences were observed in homogeneous selection among low magnetic, Earth magnetic, and high magnetic groups (52.07%, 53.08%, and 52.62%, respectively), all were significantly lower than in the unimproved control group (62.81%), suggesting that earthworms alleviated some environmental stress on substrate bacteria.

For fungal communities, drift (45.98 ± 3.59%; Fig. 6c) and homogeneous selection (15.08 ± 2.57%; Fig. 6c) played important roles, correlating primarily with substrate physicochemical properties and earthworm gut indicators, respectively (Fig. 6d). Drift was strongly influenced by substrate environmental factors (Fig. 6d), while homogeneous selection was mainly associated with earthworm gut characteristics (Fig. 6d). Although dispersal limitation was substantial (37.71 ± 5.24%; Fig. 6c), it showed weaker correlations with measured indicators (Fig. 6d). In the substrate, drift increased from 49.49% to 58.75% with increasing magnetic field strength, suggesting that fungi may experience reduced stress under high magnetic conditions, consistent with our earlier hypothesis that high magnetic fields might inhibit gizzard grinding efficiency. In the gut, the low magnetic group exhibited the highest proportion of drift and the lowest proportion of homogeneous selection, likely because earthworm digestion was so effective against fungi that the sampled gut fungi were basically recently ingested by the earthworm, which is consistent with the multivariate analysis (Fig. 4f). Although homogenizing dispersal in both bacterial and fungal communities correlated significantly with multiple earthworm gut physiological and substrate physicochemical indicators (Fig. 6b & d), its overall impact was minimal due to low proportions (1.21 ± 0.46% for bacteria; 0.51 ± 0.15% for fungi; Fig. 6a & c).

To further elucidate the regulatory mechanisms of magnetic environments on lunar soil simulant improvement, we applied partial least squares path modeling (PLS-PM) to systematically quantify multi-level interactions within the "magnetic field-earthworm-microbe-substrate" system (Fig. S5). The analysis revealed that magnetic field strength had weak direct effects on substrate physicochemical properties (path coefficient = -0.0279), primarily exerting indirect effects through strong inhibition of earthworm gut physiological function (-0.76), which in turn suppressed substrate physicochemical improvement. Specifically, impaired earthworm physiological status under high magnetic fields may both directly reduce substrate improvement efficiency and indirectly inhibit improvement by enhancing homogeneous selection in gut microbial assembly (0.88) while suppressing substrate microbial communities (-0.74). These findings suggest that earthworms may serve as central mediators in transmitting magnetic field environmental effects.

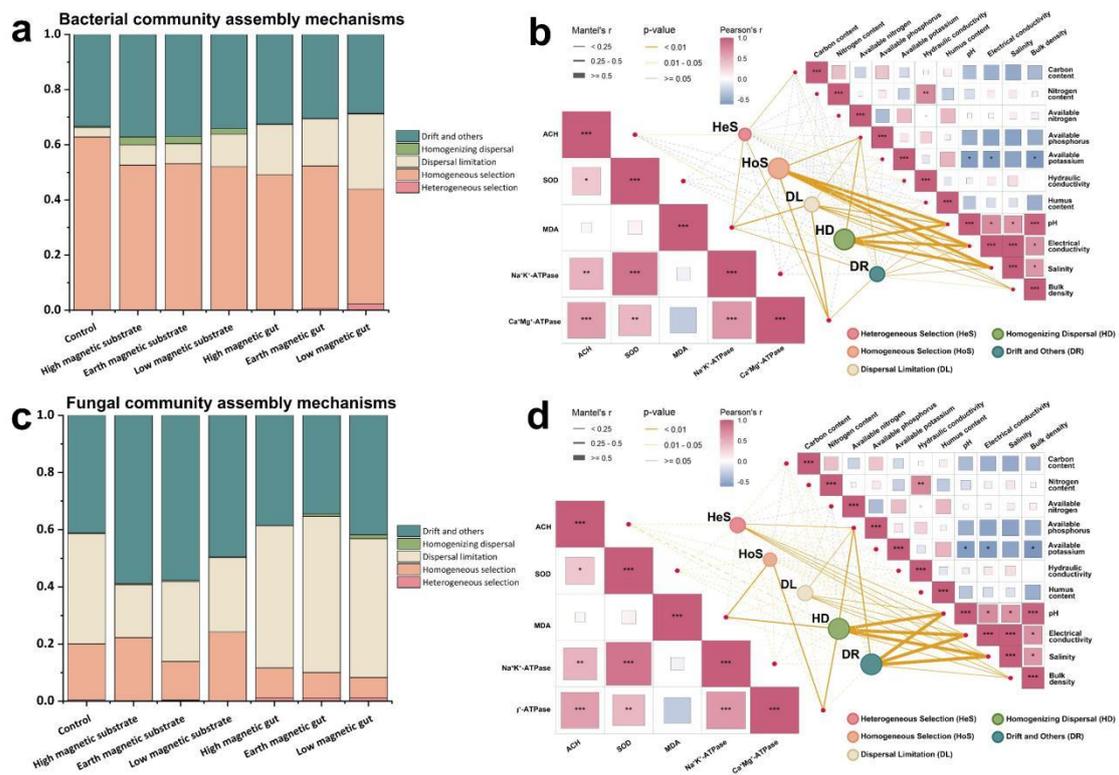

**Fig 6.** Community assembly results of substrate and earthworm gut for bacteria (a) and fungi (b), and Mantel tests between environmental/gut indicators and assembly processes for (c) bacteria and (d) fungi.

## 4. Discussion

With the deepening of research on Bioregenerative Life Support Systems (BLSS), the recycling of basic resources such as oxygen and water has been gradually achieved. However, due to the complex composition and recalcitrant nature of solid waste, its long-term closed-loop recycling has not yet been realized, becoming a key bottleneck restricting the sustainability of deep space exploration missions. Previous studies have mostly focused on improving degradation efficiency through purely physical, chemical, or biological methods, while neglecting the integrity of BLSS as an ecosystem, making it difficult to efficiently convert solid waste into renewable resources.

We found that earthworms can create a beneficial coupling between the solid waste degradation process and the improvement process of lunar soil simulant-based cultivation substrates. By synergistically degrading organic waste with gut-associated microbes, their metabolic products (such as vermicast) possess dual functions of substrate improvement and mineral nutrient recycling from waste ash. This represents a "waste-to-resource" conversion pathway that fully considers the material cycling integrity of BLSS. However, the weak magnetic environment on the lunar surface and in space may interfere with earthworm physiology and affect their degradation efficiency. Earth itself has a natural magnetic field, and all organisms live under this ground magnetic field strength (approximately 25-65 µT), while the lunar surface has an extremely weak magnetic field (5 nT) (Mitchell et al., 2008), which may potentially affect biological components like earthworms in lunar BLSS.

Therefore, this study aims to explore the interaction mechanisms of the earthworm-mediated coupling system for BLSS solid waste recycling and lunar soil simulant improvement under weak magnetic conditions, providing theoretical and technical support for closed-loop resource management in BLSS required for lunar bases and deep space exploration missions. From the perspective of research difficulty, for biological components in confined BLSS modules, compared to environmental factors such as microgravity and long-term low-dose ionizing radiation, the negative impact of weak magnetic fields is likely smaller (Council et al., 2012) but more challenging to simulate on Earth. Therefore, if it can be first demonstrated that weak magnetic fields have no negative impact on this earthworm-lunar soil improvement-solid waste recycling coupling system, it would be good news for subsequent space environment BLSS research.

In our experiments, the effects of magnetic fields on earthworms were comprehensive. First, in terms of oxidative stress, earthworms exposed to high magnetic fields showed severe oxidative stress responses (Fig. 2a), and high magnetic fields damaged earthworm intestinal tissues, while weak magnetic fields had almost no adverse effects on earthworm physiology (Fig. 2e). Some studies have found that high magnetic fields significantly increase MDA levels in yeast cells, causing severe oxidative stress while markedly inhibiting yeast cell proliferation (Kthiri et al., 2019), which shares some similarities with our results.

In terms of neurology, medicine has long linked magnetic fields to neural medicine, as static magnetic fields have significant regulatory effects on neuronal physiological functions. We observed that as static magnetic field strength increased, earthworm $Ca^{2+}$-$Mg^{2+}$-ATPase activity significantly decreased (Fig. 2b), while $Na^+$-$K^+$-ATPase activity remained unchanged (Fig. 2d). This may be because $Ca^{2+}$-$Mg^{2+}$-ATPase (such as PMCA, SERCA) functions to maintain intracellular $Ca^{2+}$ homeostasis, and its ATP hydrolysis activity is directly linked to $Ca^{2+}$ influx. Static magnetic fields can fine-tune transmembrane potential through Lorentz forces and membrane polarization effects, thereby affecting the opening probability of voltage-gated $Ca^{2+}$ channels (VGCCs), manifesting as increased channel opening probability and $Ca^{2+}$ influx under moderate field strengths. However, when field strength is too high, the orientation and rigidity of the membrane lipid bilayer and channel proteins increase, inhibiting channel conformational changes and ion penetration, leading to reduced $Ca^{2+}$ influx (Pall, 2013; Wu et al., 2022). When $Ca^{2+}$ influx decreases, the pumping demand and activity of ATPase will significantly decrease (Formaggio et al., 2025). In contrast, although $Na^+$-$K^+$-ATPase is an "electrogenic" enzyme, it does not rely on instantaneous voltage gating and is far less sensitive to fine-tuning of membrane potential than calcium channels (Bahinski et al., 1988). Additionally, $Na^+/K^+$-ATPase often coexists in various cell types as different isoforms (α1-α4), and its activity is dynamically regulated by multiple factors such as intracellular $Na^+/K^+$ concentration and kinase phosphorylation states (Li and Langhans, 2015), making it difficult to be disrupted by weak external magnetic fields.

Furthermore, acetylcholine levels in earthworms were significantly reduced under high magnetic fields (Fig. 2c), possibly because the release of neurotransmitters such as acetylcholine is triggered by calcium ions (flowing within the presynaptic membrane of nerve terminals) (Stanley, 1993), and when calcium ions are interfered with by magnetic fields, acetylcholine levels may decrease. In terms of muscles, high magnetic fields can disrupt F-actin fibers in bovine adrenal chromaffin cells (Ikehara et al., 2010), affecting cell structure and movement, which may be the mechanistic reason for earthworm muscle cell damage under high magnetic fields (Fig. 2e).

The physicochemical data of improved substrates can indirectly reflect the effects of magnetic fields on earthworm ecological functions. In this study, substrates under high magnetic fields performed poorly in multiple key indicators (available nitrogen, available phosphorus, humus) (Fig. 3) and had relatively higher pH, conductivity, and salinity (Fig. 3 & Fig. S1), which are unfavorable for most plant growth (Fig. S1e). We know that muscles and nerves are closely related, and magnetic fields may alter the activation and inactivation states of $Ca^{2+}/Na^+$ channels(Picazo et al., 1995), thereby changing the efficiency of neural signal transmission and muscle contraction ability, further affecting the contraction patterns of earthworm intestinal and gizzard muscles. This may reduce food residence time in the intestines, decrease grinding efficiency, alter food physical state, and lead to incomplete digestion. Additionally, earthworms exposed to magnetic fields exhibit behavioral deviations, potentially affecting their environmental perception, including food recognition and localization (YALÇIN et al., 2020). These findings suggest that magnetic fields may interfere with earthworm behavior and physiology by affecting neuronal electrical activity or muscle cell contraction ability, weakening earthworm feeding and digestion capacities, thereby impacting substrate improvement.

Moreover, magnetic fields may also affect substrate improvement by influencing the metabolic activity, survival rate, and distribution of substrate or earthworm gut microbes. Some studies have shown that Escherichia coli cultured in magnetic fields above 480 μT exhibit faster growth rates than controls (Aarholt et al., 1981). Magnetic fields can also regulate bacterial movement behavior and hydrodynamic properties, further affecting their distribution and interactions in the gut. For example, weak uniform external magnetic fields (~3 mT) and local micro-magnetic fields can adjust bacterial orientation, thereby controlling their swimming behavior (Mumper et al., 2017). In our study, the impact of magnetic fields on the overall species abundance of environmental microbial communities was not significant (Fig. 4), possibly because our defined "high magnetic field (70 μT)" was relatively weak compared to industry studies, limiting result comparability. However, some trends were still observable. For instance, the fungal Shannon index in substrates under high magnetic fields was significantly higher than under Earth and weak magnetic fields (Fig. 4d), possibly because high magnetic fields weakened earthworm feeding and fungal digestion abilities (mentioned above), thereby reducing suppression of dominant fungal species.

Compared to magnetic fields, earthworms had more significant effects on microbial communities. The Chao1 and Shannon indices of earthworm gut bacteria were significantly higher than those of control substrates and further increased after being excreted back into the substrate (Fig. 4a & b), demonstrating earthworm promotion of substrate bacteria. The high humidity, neutral pH, and abundant organic matter in earthworm guts can provide suitable habitats for specific microbes, such as degraders like *Proteobacteria* (e.g., *Pseudomonas*) and *Actinobacteria* (Sapkota et al., 2020), which efficiently degrade complex organic matter like cellulose and proteins. In contrast, earthworms primarily suppressed fungal diversity (Fig. 4d), leaving only fungi adapted to earthworm treatment; gut fungal diversity was largely consistent with substrate fungi (Fig. 4d), suggesting gut fungi may mainly originate from substrates with no native fungi. Earthworms can destroy fungal hyphae by secreting chitinase (the main component of fungal cell walls is chitin) and degrade hyphae through digestive grinding, also killing fungal spores (Dempsey et al., 2011).

The molecular ecological network structure of microbial communities can reveal the influence of environmental factors on microbial interactions (Zhou et al., 2010), where modules represent independent functional units (Deng et al., 2012a). In terms of the proportion of positive correlations, microbial networks in weak magnetic substrates and Earth magnetic substrates both showed approximately 49% positive connections (Fig. S4), indicating similar shaping effects on substrate microbial communities and suggesting weak magnetic fields had minimal impact on microbial function and structure. In contrast, networks in high magnetic substrates and unimproved substrates both exhibited around 54% positive correlations (Fig. S4), implying that high magnetic fields may inhibit the improvement process at the microbial community level, making microbial interactions more similar to the unimproved state. For microbial communities, lignocellulosic plant residues represent an abundant survival resource (albeit recalcitrant), and their degradation is a cooperative "task" (Mizrahi et al., 2021). This suggests that during the maturation of soil involving lignocellulosic residue degradation, a higher proportion of positive correlations in microbial networks may indicate ongoing lignocellulose degradation, with incomplete substrate maturation and improvement. Therefore, high magnetic fields likely reduce substrate improvement efficiency (Fig. 3) by inhibiting lignocellulose degradation and maturation processes, as evidenced by the elevated proportion of positive correlations in their networks (Fig. S4).

Regarding the earthworm gut microbial networks, it is readily observable that Network Size, Average Path Length, and Modularity all progressively decreased with increasing magnetic field strength (Table S1). This suggests that magnetic fields exerted certain inhibitory effects on microbial interactions, which may partially explain the reduced substrate improvement efficiency under high magnetic conditions (Fig. 3). We have previously reported similar findings where external stress (simulated microgravity) strongly suppressed network interactions in lignocellulose-degrading microbial communities (Liao et al., 2025). The proportion of positive correlations in gut networks still showed a marked increase with field strength (Fig. 5), again indicating slower lignocellulose degradation and incomplete maturation processes. Compared to substrates, microbial networks in earthworm guts exhibited more pronounced shaping by magnetic fields (Table S1, Fig. 5), likely because magnetic effects on animals are more direct and intense, with these impacts being manifested through microbial networks. Ultimately, weak magnetic fields significantly enhanced both the network scale and functional redundancy of earthworm gut microbiota.

Community assembly results further corroborated these network analysis conclusions. The data revealed that under weak magnetic fields, both bacterial and fungal communities in earthworm guts showed the lowest proportions of deterministic processes (primarily homogeneous selection) among all treatment groups (Fig. 6a & c), indicating relatively lower environmental stress. In such conditions, microorganisms no longer required tight cooperation for survival, potentially leading to reduced synergistic interactions (Yang et al., 2024) (Fig. 5 & S4). Notably, in analogous composting systems, such decreased synergy has been associated with lower environmental stress (Zhao et al., 2023) and more complex community functionality (Chen et al., 2022). Consequently, weak magnetic substrates may ultimately develop into microecosystems characterized by functional diversity and reduced environmental pressure (Zhou and Ning, 2017). Additionally, within substrates, the decreased fungal drift under weak magnetic fields suggested increased environmental stress on fungi (Fig. 6b & d), demonstrating that weak magnetic fields inhibit fungal colonization in substrates. This inhibitory effect may prove beneficial for plant growth and the enhancement of ecosystem functions (Wardle et al., 2004).

## 5. Conclusion

This study reveals the effects of different magnetic field strengths on earthworm physiology and the structure and function of environmental microbial communities. By simulating the weak magnetic environment of the lunar surface, we found that earthworms under weak magnetic conditions can effectively maintain their physiological health and significantly improve substrate physicochemical properties. These improvements include reducing substrate alkalinity and salinity, increasing available nutrient content, and enhancing microbial diversity and functional complexity.

This is of great significance for establishing BLSS in lunar bases. Furthermore, microbial network analysis showed that under weak magnetic conditions, the modularity and robustness of microbial communities significantly increased, helping to enhance ecosystem stability. Therefore, future research could further explore the long-term evolution of microbial communities under different magnetic conditions and assess the potential impacts of these changes on ecosystem function. Additionally, understanding the roles of key nodes in microbial networks and their performance in different magnetic environments will aid in developing more robust ecosystem management and restoration strategies.

This study not only advances our understanding of biological treatment components in BLSS but also provides new strategies and theoretical foundations for solid waste treatment and ecological function restoration in future deep space missions, laying the groundwork for establishing sustainable ecosystems on the Moon or other planets.

## CRediT authorship contribution statement

Sihan Hou & Zhongfu Wang: Writing – original draft, Methodology, Visualization, Analysis. Yuting Zhu: Visualization. Jiajie Feng: Writing – review & editing, Funding acquisition. Hong Liu: Supervision, Funding acquisition. All authors read and approved the final manuscript.

**Declaration of competing interest**

The authors declare that they have no known competing financial interests or personal relationships that could have appeared to influence the work reported in this paper.

**Acknowledgments**

This work was financially supported by the grant from the National Natural Science Foundation of China (32200104) and the Fundamental Research Funds for the Central Universities. We thank Hui Liu and Jingkai Tang for their their design of the permalloy shielding device and provision of magnetic field parameters, Yuming Fu for his suggestions on the statistics, and Letian Chen for his help on visualization.

**Data availability**

Data will be made available on request.